\documentclass[aps,twocolumn,showpacs,preprintnumbers,amsmath,amssymb,floatfix]{revtex4}
\usepackage{graphicx}
\usepackage{epsfig}
\begin{document}
\def\be{\begin{equation}}
\def\ee{\end{equation}}
\def\bearr{\begin{eqnarray}}
\def\eearr{\end{eqnarray}}
\def\tc{$T_c~$}
\def\half{$\rm \frac{1}{2}$~}
\def\Moire{$\rm Moir\'e $~}

\title{Theory of Emergent Josephson Lattice in Neutral Twisted Bilayer Graphene\\
(Moir\'e is Different)}

\author{G. Baskaran}

\affiliation
{The Institute of Mathematical Sciences, C.I.T. Campus, Chennai 600 113, India \&\\
Perimeter Institute for Theoretical Physics, Waterloo, ON N2L 2Y5, Canada}

\begin{abstract}
	
`More is Different' \cite{PWAMoreIsDifferent} in graphene. A bilayer and a twist spring surprises. Recently discovered superconductivity (Tc$\approx$ 1.7 K) at an ultra low doping density $\sim 10^{11}/cm^2$ has alerted the community to look for an electron-electron interaction based mechanism, as phonon-induced attraction seems inadequate. We suggest a mechanism of superconductivity, where an important role is played by the dense (density $\approx 2 \times 10^{15}/cm^2$)  \textit{$\pi$-electron fluid of graphene layers}. This fluid bears off-shell resonating valence bond correlations (RVB) at the carbon-carbon bond scale. A commensurate twist $\theta \approx 1.1^o$, creates charge neutral carrier puddles (size $\sim$ 50 \AA) and forms a triangular Moir\'e lattice of local AA registry. AA registry dopes equal numbers of electrons and holes via interlayer tunneling, whereas AB registry does not. Carriers inside the charge neutral puddles form equal numbers of -2e and +2e Cooper pairs, using on-shell RVB correlations. A Josephson-Moir\'e lattice emerges. Coulomb blockade competes with pair tunneling and creates a Bose Mott insulator. Gate doping dopes the Bose Hubbard model and creates superconductivity. Our message is that RVB correlations, which remain dormant in (carrierless) neutral graphene become on-shell for two added electrons, as they are indistinguishable from electrons that make the background $\pi$-fluid in graphene.

\end{abstract} 
\maketitle

\section{Introduction}
Emergence and \textit{quantum complexity} abound in the world of graphene \cite{CastroNetoRMP,KatsnelsonBook,FrancoNoriReview,GBQComplexity}. A twisted bilayer, containing two graphene layers, become \textit{more than sum of the parts} \cite{Cao1,Cao2},
via Moir\'e lattice modulated electronic properties. Single electron spectrum changes in a remarkable fashion \cite{CastroNeto2007,Mele2010,flatband1,flatband2,flatband3,PacoNonAbelian2012,topologicalBandKindermann,MoirePicture,ShasyNanda}. Uniform magnetic field creates a  Hofstadter butterfly fractal spectrum \cite{vonKlitzing,McDonaldBFly,MeleComb,PhilKimBFly,BrokenSym}
This in turn is capable of producing interesting manybody effects and phases  with and without magnetic fields. Quantum Complexity and emergence continues to grow in graphene.

In a series of recent exciting experiments, prisine twisted bilayer graphene has been shown to behave as an insulator\cite{Cao1}, which becomes a superconductor \cite{Cao2} on gate doping. It has been suggested that the twist isolates a low density of electron carriers at the Fermi level and builds a Fermionic Mott insulator subsytem, which provides a template for the experimentally observed low Tc superconductivity. Twisted trilayer graphene with ABC stacking grown on a hexagonal boron nitride layer also shows \cite{WatanableTrilayer} interesting gate tunable metal insulator transitions. 

The aim of the present article is to present microscopic and qualitative considerations for twisted bilayer graphene, and bring out certain inescapable, general and novel features. We present a commensurate twist ($\theta \approx 1.1^{\circ}$) induced insulating Josephson-Moir\'e lattice of Cooper pair puddles and suggest a Boson Mott insulator model. Gating dopes the Boson Mott insulator and creates superconductivity. 

Recent experimental results \cite{Cao1,Cao2} have inspired a lot of interesting theoretical work and ideas \cite{strainTrambly,AllanMcDonald18.2,Pal,LeonBalents,ManishJain,LiangFu,AshwinSenthil,BitanRoy}. A general emphasis is on  emergent Fermionic Mott insulators. Theoretical results also provide important insights into existence of few nearly flat bands close to the Fermi level for small twist angle, in the background of a \textit{complex band structur that is sensitive to twist angle}. Our work brings out a new local aspect, namely Cooper pair formation within a Moir\'e supercell, using properties of the nearly flat bands and point to a new direction.

We briefly intoduce our idea first. In an AA stacked bilayer, a case of perfect registry of atoms, interlayer tunnelling creates small electron and hole Fermi pockets of equal area. AB stacking on the other hand, does not ceate any Fermi poclets; bilayer remains a semimetal, with a qudratic band touching. A commensurate angular twist from AA registry creates a triangular Moir\'e superlattice of local AA registry. This superlattice contains \textit{charge neutral puddles} of quantum confined electron and hole carriers, generated via local interlayer dynamics. Partial registry in rest of the sample, a honeycomb superlattice containing AB or BA local stacking, does not acquire carriers. 

Quantum confined states close to Fermi level weakly overlap and form few nearly flat bands. This also follows from band theory. Carriers in the narrow bands have to get paired in order to create superconductivity. \textit{Phonon induced pairing among the few carriers, because of quantum confinement, is negligibly small}. Pairing between two electrons in neighboring Moir\'e supercells is also heavily supressed in view of the large supercell lattice parameter $\sim$ 130 \AA, for $\theta \approx 1.1^0$. 

The required pairing is provided by the dense background $\pi$ electron fluid of graphene substrate, via resonating valence bond correlations it nurtures. As carriers are absent at Fermi level in graphene, RVB correlations remain off shell and are inconsequential at low energies. However, added free carriers  bring RVB pairing on shell, as they are indistinguishable from the electrons of background dense $\pi$ electron fluid. Cooper pairs get formed in charge neutral puddles. A Josephson lattice is formed. Coulomb force compete with Cooper pair tunneling. We get a Cooper pair Mott insulator. External gate voltage dopes the Boson Mott insulator and control superconductivity.

Our paper is organized as follows. We first discuss presence of self doping in the case of perfect AA stacked bilayers and absence of self doping in the case of AB stacked bilayers.  Then we study twisted bilayer graphene at the small magic commensurate angle and discuss the origin of pairing. We present an emergent Josephson lattice in a Mott insulating state. A doped Bose Hubbard model is suggested as a minimal model to understand the experimentally seen insulator to superconductor transition. 

In the appendix we summarise earlier attempts to understand electron correlation effects in the broad band graphene and RVB idea based prediction of high Tc superconductivity in graphitic systems.\\

{\bf RVB superconductivity in graphene based Systems.}
Before we begin, a remark about doped graphene and graphite is in order. In RVB theory (explained in the supplement) one expects superconductivity at room temperature scales for a carrier doping of $\approx 3 \times 10^{14} /cm^3$). This density is about 15 \% of the graphene $\pi$ electrons. It is indeed interesting that historically there have been intriguing signals \cite{Kopelevich,Esquinazi,Volovik1} for \textit{elusive superconductors}, even at room temperature scales ! The new experiments \cite{Cao1,Cao2}, in view of the nearly 100 times lower carrier density is not inconsistent with RVB mechanism of superconductivity at 1.7 K. Are we witnessing tip of an iceberg in the recent experiments \cite{Cao1,Cao2} ?

In an important work Choi and McKinnon \cite{ChoiMckinnon} studied the possibility of high Tc superconductivity in intercalated graphite from RVB physics point of view.
The physics of their work is very similar to the one we discuss in the supplement.
Even though intercalated graphites have very high doping density, ordered intercalant atom encourages a strong charge ordering. This ordering competes with RVB mechanism and reduces Tc significantly to the observed small values \cite{GB1}.

\section{Self Doping in AA Stacking and its Absence in AB Stacking}

In this section we review known results about how interlayer tunnelling modifies the band structure close to the Fermi level in the case of AA and AB stacking. AA stacked graphene has been studied in the past from experimental \cite{AAExpt1,AAExpt2} 
and many body theory \cite{FrancoNoriAAAFM,FertigAABilayer,SahuAAAFM} points of view.

Consider the single particle Hamiltonian of an AA stacked bilyer graphene. As AA stacking creates a perfect registry between atoms in the two layers, interlayer hopping has full translational invariance of the underlying honeycomb lattice. The largest interlayer hopping t$_0$ is between two atoms, aligned along c-axis. Band theory estimates a value t$_0 \sim$ 0.35 eV. 

\begin{figure}
\includegraphics[width=0.5\textwidth]{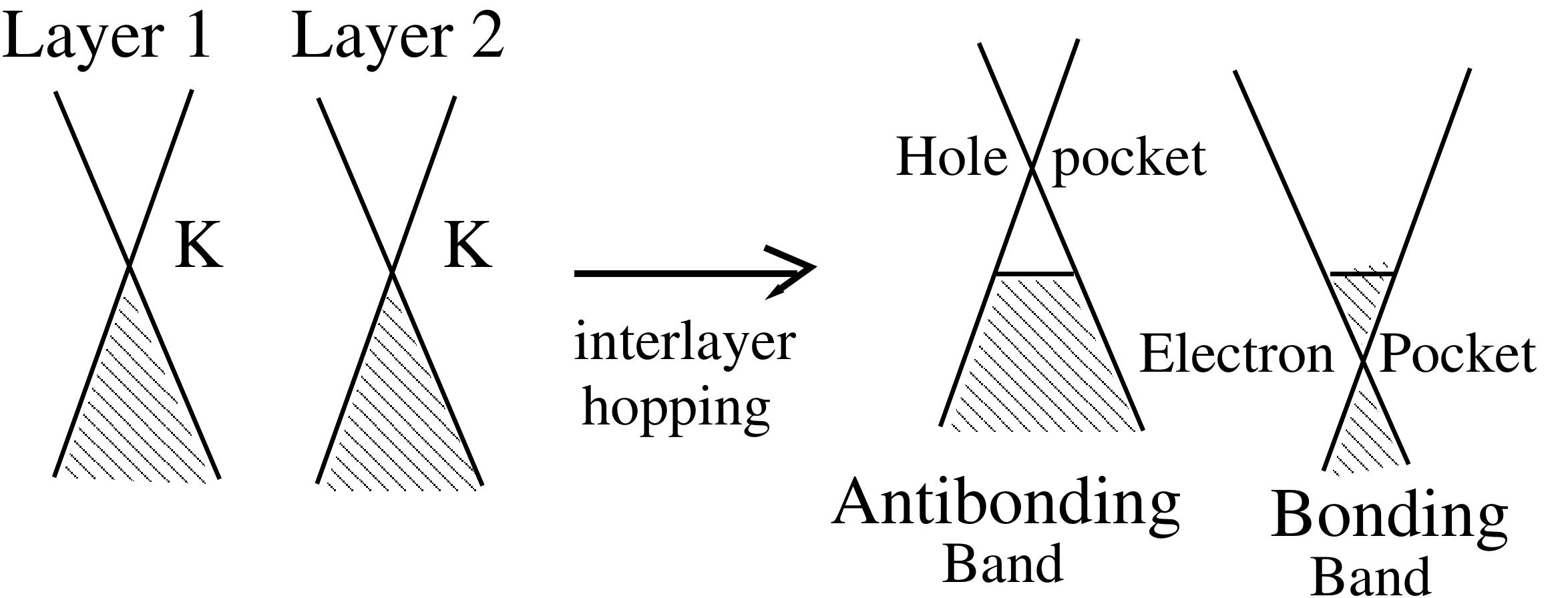}
\caption{Dirac semimetallic states at K point of two layers in the AA stacked bilayer get hybridized by interlayer hopping t$_0$. Resulting bonding and antibonding interlayer bands rigidly shift in energy in opposite directions. It results in creation of electron like and hole like Fermi pockets of identical area with Fermi energy t$_0 \approx$ 0.35 eV.} \label{Fugure 1}
\end{figure}

By symmetry two identical Bloch states from top and bottom graphene layers alone hybridize and split into (interlayer) bonding and antibonding states. Bonding and antibonding bands thus formed rigidly move up and down by an energy t$_0$. This creates overlapping electron and hole Fermi pockets of identical sizes (ignoring a small trigonal warping) at K and K' points in the BZ (Figure 1). 

Thus interlayer dynamics adds an equal density of electron and hole carriers to neutral graphene bilayer. Fermi momentum of the pockets is given by $\dfrac{2t_0}{3 t a}$, where t ($\approx$ 3 eV) is the nearest neighbor hopping in a graphene layer and a is the nearest carbon carbon distance in a layer. Aerial density of added carriers is $\sim 4 \times 10^{12} /cm^2$. This is very small and is about 0.2 \% of density of p$_z$-electrons $\sim 2 \times 10^{15}$ (or carbon atoms) in neutral graphene layer. 

In the case of AB or BA stacked bilayer, perfect registry exists only between a pair of triangular sublattice, one from the top and other from bottom layer. An interlayer hopping t$_0 \approx$ 0.35 eV exists between half of the atoms that are in registry. Remaining triangular sublattices, which are out of registry, have negligible interlayer hopping matrix elements. Consequently AB stacking maintains semi metallicity, even after interlayer hybridization; no Fermi pockets are produced. There are two important differences however. Of the two interlayer hybridized bands, i) one has, instead of two Dirac cones, two touching quadratic bands at K and K' points at the Fermi level and ii) the other acquires a gap ($\sim$ t$_0$) at the Fermi level.

The above known facts form the basis of our proposal, discussed in the next section, for the observed low Tc superconductivity and insulating behavour seen in gated experiments in twisted bilayer graphene, very recently.
 
\section{Twisted Bilayer Graphene and Emergent Josephson Lattice}

Even at the level of free  electron tight binding model, a twisted bilayer graphene offers challenge for analytic calculation, A space dependent modulation of interlayer hopping hybridizes Bloch states of two layer, that are all related by reciprocal lattice vectors of the Moir\'e lattice. Corresponding single particle diagonalization problem can become unwieldy. Physics motivated accurate analytical approaches have been made however \cite{CastroNeto2007,Mele2010,flatband1,flatband2,flatband3,PacoNonAbelian2012,topologicalBandKindermann,MoirePicture,ShasyNanda} 

There has been interesting experimental observations of Moir\'e patterns and experiments in graphene \cite{Moire1,Moire2,STMTwist,DynamicDecoupliing,CoherenceBreadown}. Very recent 
works have also performed interesting experiments on various aspects of twisted bilayer graphene\cite{MacDonaldExptPNAS2007,Arindam,Interface2018,transportThroughTop}.
An interesting recent theoretical work has anticipated magnetic states \cite{PacoTheory2017} in twisted bilayer graphene at small twist angles $\theta \approx 1^0$, before the wave of current experiments.

In what follows we identify certain general and robust feature at smal commensurate magic angles and introduce a model and estimate parameters of the model.\\ 

\begin{figure}
\includegraphics[width=0.5\textwidth]{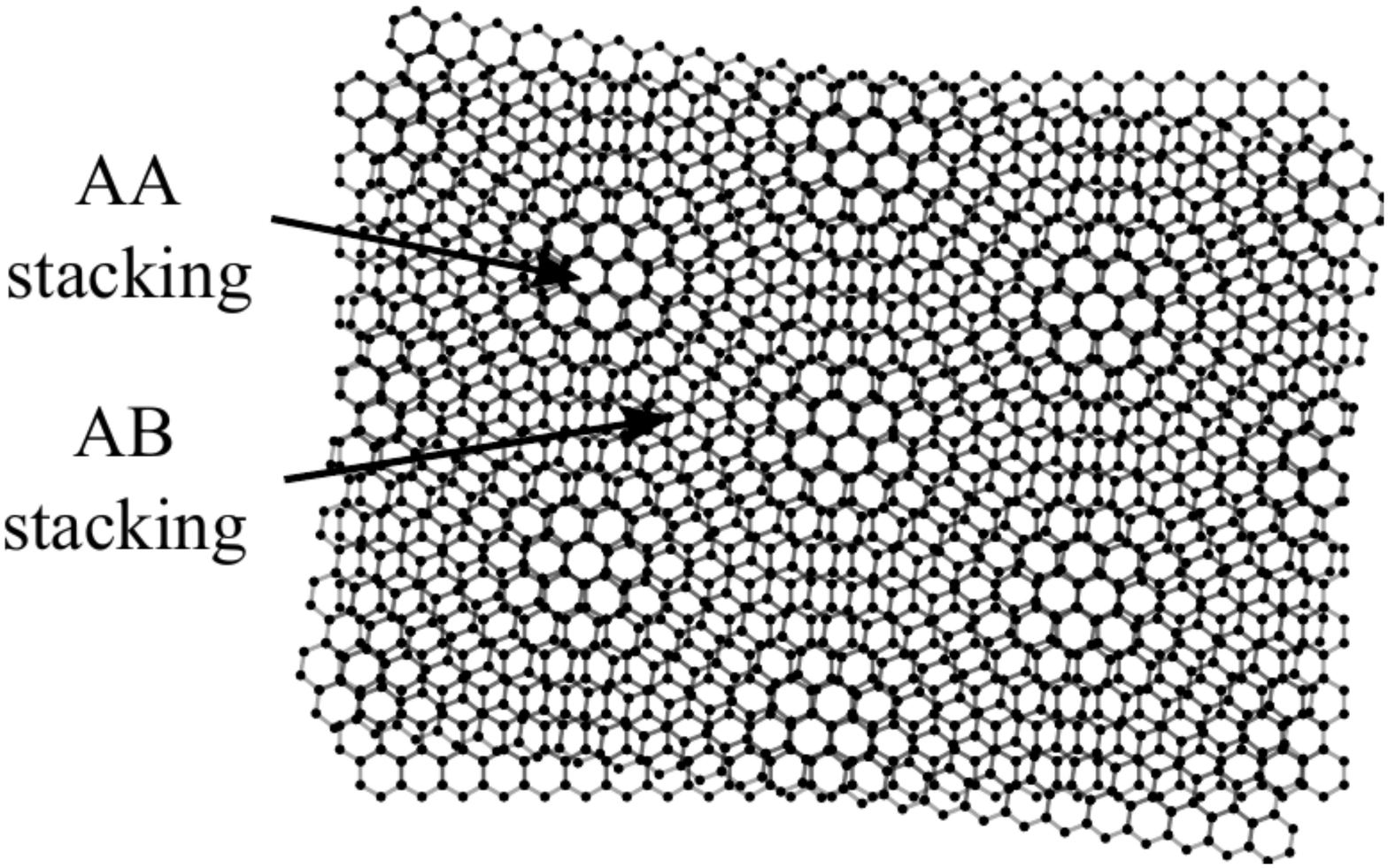}
\caption{A Moire pattern for a small commensurate angle twist in bilayer graphene (Acknowledgement: this figure is taken from reference \cite{strainTrambly}). Within a region of dominant AA stacking, interlayer tunneling add equal number of electron and hole carriers and quantum confines them. In our theory they become Josephson islands, containing RVB correlation induced pairing and equal number of $\pm$ 2e Cooper pairs.} \label{Figure 2}
\end{figure}

\textbf{Number of Carriers in a Puddle.}

The Moir\'e superlattice (Figure 2) has a three (triangular) sublattice structure. These sublattices contain dominant bilayer registries of AA, AB and BA types respectively.We focuss on the experimentally relevant small magic twist, $\theta \approx 1.1^0$. Unit cell size of the superlattice is large $\approx$ 130 \AA; it contains $\sim$ 6000 atoms \cite{MoirePicture}, As AA registry alone adds electron and hole carriers via interlayer tunneling, we get a periodic Moir\'e lattice of charge neutral puddles. In a semiclassical sense, we say that in the Moir\'e lattice \textit{a periodic modulation of the Fermi wave vector in real space occurs.}

Number of carriers in the puddle is estimated as follows. Local AA stacking in a Moir\'e supercells produces the puddle. Doping density for AA stacking is about 0.2 \% of the aerial density of carbon atoms in a graphene layer. In the Moir\'e supercell containing 6000 atoms, about a third, 2000 atoms, have approximate AA registry. Thus a rough estimate of added carrier number is, 0.2 \% of 2000, which is about 4 per puddle. This estimate is in the right ball park, compared to estimates from band structure results.
 
\textbf{Origin of Nearly Flat Bands.} A modulated interlayer dynamics not only adds carriers but also produces quantum confinement. Depth of the \textit{confining pseudo potential} is determined by the maximum value t$_0\approx$  =.35 eV, of the modulated interlayer hopping matrix element.

Physically,  Moir\'e superlatttice band splitting scale is given by
\be
\Delta E \sim h v_F G_{SL},
\ee
where v$_F$ is Fermi velocity in graphene layer and G$_{SL}$ is the reciprocal lattice vector of the Moir\'e superlattic. For our supercell, level splitting is  $\Delta E \sim 0.1 eV $. This splitting is small compared to Fermi energy of added carriers, t$_0 \approx 0.35$ eV. This is consistent with band theory result, where one sees few nearly flat bands in the energy range of 0.35 eV (Fermi energy of the untwisted AA bilayer).

A closer look at the band structure reveals that while details of band profile change sharply with twist angle, some general features survive. For example, for the case close to $\theta$ = 1$^0$, we have two to three flat bands at the Fermi level in the scale of the interlayer tunneling matrix element. As we mentioned ealier, these narrow bands are derived from states of the small Fermi pocket Bloch states of electrons and holes. Thus it is natural that a few electrons and holes, in equal numbers, live in these states and get scattered among these narrow band states from interaction processes. 

\textbf{Quantum Confinement and Narrow Bands}. Quantum confined single particle states between neighboring cells overlap weakly and form very narrow bands. We estimate an upper bound of this as follows. Quantum confined eigen functions are spread over $\sim$ 2000 atoms. Hopping between states from neighboring cells arise, via nearest neighbor single particle hopping t ($\approx 3 eV$) between about 20 to 30 atom pairs at the overlapping boundary region. Further, amplitude of confined wave functions sharply fall at the overlap region C, $\psi(C) \sim \frac{1}{\sqrt{10}}$. This gives us  hopping between neighboring supercells, 
\be
t_s \sim \frac{1}{(\sqrt{2000})^2} \times
\frac{1}{10} \times 20 \times t \sim  1~\rm meV. 
\ee
This estimate is in the right ball park and consistent with width of narrow bands as given by band structure results.

Under some conditions one may obtain flat bands, arising from special interference effects and symmetries. However, unavoidable perturbations are likely to broaden these bands. 

\textbf{Origin of Pairing and Scale of Cooper Pair Binding Energy}

So far we have seen that neutral carrier puddles containing equal and a small  number of electrons and holes are present in each Moire unit cell, close to the Fermi energy. As the confinement energy scales are large, a phonon mediated attraction can not lead to requred pairing strength \cite{Cao1,Cao2}.

How about inter Moir\'e supercell singlet pairing in an emergent Fermion Mott insulator ? As single particle tunneling matrix elements between neighboring supercells is very small our estimates show that such processes are also strongy inhibited. This reduces probability of intersupercell Cooper pairing, via superexchange for example.

We find that RVB correlation, which is present, but is of no consequence in neutral graphene, because of absence of carriers at Fermi level, could help in the following fashion. That is RVB correlations modify the ground state, but keep low energy physics the same as a Fermi liquid semimetal and do not produce superconductivity. As explained in the supplementary sections a moderate Hubbard U in the broad band graphene nurtures a singlet correlation. This attraction is off-shell for electrons that make up netural graphene. Two added free carriers, which become indistinguishable from $\pi$ electrons that make up graphene, also feels the same RVB correlation, but onshell.

In our early theory for graphite like systems we incorporated the above (Pauling's RVB physics) via a phenomenological pairing Hamiltonian added to the free electron Hamiltonian: 
\begin{equation}
H_{\rm RVB} = - t \sum_{\langle ij \sigma \rangle} (c^\dagger_{i\sigma}c^{}_{j\sigma} + H.c.) - J \sum_{\langle ij \rangle} b^\dagger_{ij}b^{}_{ij}
\end{equation}
Here, $b^\dagger_{ij} \equiv
\frac{1}{\sqrt{2}} (c^\dagger_{i\uparrow}c^\dagger_{j\downarrow} - c^\dagger_{i\downarrow} c^\dagger_{j\uparrow})$ is the singlet electron pair operator.  The meaning of the above term is the following. Each head on collisions involving a double occupancy of opposite spin electrons (spin singlet state) lead to a quantum entanglement in the spin singlet (not in the triplet) channel between two electrons. This spin entanglement is strongest when they are neighbors. A large pairing pseudopotential J, in our model Hamiltonian forces such short distance quantum entanglement. As discussed in the supplement, this large pseudopotential for graphene is J $\sim$ 2 eV. This large J value is also supported by our spin-1 collective mode theory of graphene.

The above short distance attraction term scatters charge -2e electron pairs and charge +2e hole pairs in the puddle among available quantum confined levels close to the Fermi level and creates a kind of \textit{zero momentum} state within a puddle for charge $\pm$ 2e Cooper pairs. A lower bound for binding energy of Cooper pairs is obtained as an energy gain between two electrons, when they happen to be nearest neighbors in the graphene lattice, during their sojourn in quantum confined states. As probabiity of two pairs being neighbors in a given quantum confined state is  $\frac{1}{2000}$,  effective Cooper pair binding energy scale is
\be
\Delta_c \sim \frac{J}{2000} \approx 1~{\rm meV}. 
\ee
This scale is also related to a spin gap scale within the large island containing paired carriers; i.e., energy needed to convert a singlet pair into a triplet.

Number of Cooper pairs obtained by the above for our case of $\theta \approx 1.1^0$ is a few. Electrons and holes also will experience an attraction within the puddle. However, their singlet binding scale is reduced by the fact one lives in a interlayer bonding state and the other in the bonding state. Further, semimeteallic screening from AB and BA stacking region reduce the attraction. It is safe to say that we will be left with a couple of equal number of charge $\pm$ 2e Cooper pairs.

\textbf{Cooper Pair Tunneling Matrix Element}. Charged Cooper pairs (Bosons) undergo Josephson or pair tunnelling between two neighboring supercells. There are various processes possible, including Coulomb interaction assisted pair tunneling. We focus on pair tunneling aided by single particle tunneling t$_s$ (equation 3), via a second order single electron tunneling process, involving a breaking of Cooper pair (equation 4) and recombination. Cooper pair hopping matrix element is 
\be
t_b \sim \frac{t_s^2}{\Delta_c} \sim ~1~ \rm meV. 
\ee

\section{Cooper Pairs and Bose Hubbard Model}

From the above discussion and estimates of parameters we conclude that at low energy scales of interest, Cooper pair degrees of freedom in each cell is more relevant than added single electron or hole degree of freedom. As a first approximation we treat each supercell as a Cooper pair box, containing Cooper pairs in certain effective lowest (analogue of zero momentum) orbital state. We model this by introducing just two ($\pm$) Bose oscilators at every cell, corresponding to $\pm$ 2e Cooper pairs in the i-th supercell. 
A state containiing n Cooper pairs corresponds to n-th excited state of the Bose oscillator.  We represent Boson creation and annihilation operators by $(b^\dagger_{i\pm}, b^{}_{i\pm} ) $ and number operators by n$_{i\pm}$. 

In the absence of an external gate voltage every supercell of our twisted bilayere graphene is charge neutral. Each cell contains equal numbers of charge $\pm$ 2e Cooper pairs. Coulomb forces try to maintain charge neutrality within a cell by a bosonic Hubbard repulsion U$_{\rm B}$, given by 
\be
U_{\rm B} \sim \frac{e^2}{4\pi\epsilon_0 l_s} \approx 10~{\rm meV}
\ee
where $\epsilon_0$ is the local dielectric constant in the quantum confined region
and l$_s$ is the Moir\'e supercell lattice parameter.

Two chemical potentials $\mu_{\rm B \pm}$ fix total number of charge $\pm$ 2e Cooper pairs. In the absence of a gate voltage, for charge neutral twisted bilayer $\mu_{\rm B -}= \mu_{\rm B +}$.

Resulting simplified Bose Hubbard model for our Josephson lattice is:
\bearr
H_{\rm B} &=& -t_{\rm B} \sum_{\langle ij \rangle} (b^\dagger_{i\pm}b^{}_{i\pm} + H.c.) + U_{\rm B} \sum_i (n_{i +} - n_{i -})^2 \nonumber \\
&+& \mu_{\rm B -}\sum_i n_{i -} + \mu_{\rm B +}\sum_i n_{i +}
\eearr

Our estimate of U$_{\rm B} \approx 10~{\rm meV}$ is larger than the boson band width $\sim 6$ meV. 
Charge neutrality demands that averaage number of holes and electrons are equal in every puddle. 
Thus, in the absence of a gate voltage, $\mu_{\rm B -}= \mu_{\rm B +}$. A overwhelming Coulomb repulsion localizes bosons and we have a Boson Mott insulator, in the absence of gating.

A gate voltage adds a finite mean charge density, either electrons or holes, to the quantum confined regions.
Using the same local RVB correlations, these carriers tend to get singlet paired within the cells. Effectively, gating adds either charge +2e Cooper pairs or charge -2e Cooper pairs. This changes the local Boson occupation from a commensurate integer value. Thus we get a doped Boson Hubbard model and hence superfluidity (superconductivity) of charged Boson Cooper pairs.

Aim of the present article is to setup a correct model and discuss general physical consequences of the model. We hope to discuss our Boson Hubbard model, in the light of recent experimental results in the future.

\section{On Recent Works}

Recent experimental \cite{Cao1,Cao2} papers have suggested possible role of emergent Mott insulator in the small angle twisted bilayer graphene. While finishing this manuscript, we found some interesting theoretical articles \cite{strainTrambly,AllanMcDonald18.2,Pal,LeonBalents,Volovik1,ManishJain,LiangFu,AshwinSenthil,BitanRoy} on various aspets of twisted bilayers.

Some of them focus on the emergent fermionic Mott insulator aspect.
Xu and Balents \cite{LeonBalents} introduce a two orbital Hubbard model in a triangular lattice and discuss physics of doped Mott insulator. Po, Zou, Viswanath and Senthil \cite{AshwinSenthil} and Liang \cite{LiangFu} introduce a honeycomb lattice fermionic Mott insulator. Cooper pairs in these models are intersite and superexchange in origin. Further they invoke coulomb interactions at the length scale of Moir\'e lattice spacing
(130 \AA), for superexchange pairing for example. The background dense fluid of $\pi$ electrons does not play any role in paiiring.

In our work we find a substantial pairing in the triangular lattice of puddles. In other words what we have is an intracell pairing. Further our pairing arises from the local U, contained in the physics of the background dense $\pi$ electron fluid.

In our estimates the emergent fermionic Mott insulators are somewhat fragile compared to relatively robust Bosonic Mott insulator we have suggested. More work is needed to confirm this and also understand possible synergies. 

\section{Discussion}

Anderson's `More is different', a general notion of emergence is relevant for the broad field of condensed matter physics, biology, science and beyond. It is gratifying that `More is different' finds a place even in an apparently small world of graphene, a single elemental solid, via an emergent hierarchy, quantum complexity \cite{GBQComplexity} and surprises. One is also reminded of elemental liquid He3 \cite{VolovikBook} and elemental solid bismuth \cite{GBBismuth}.

Authors of recent experiment in twisted bilayer graphene correctly point out that \cite{Cao2}, a very low carrier density rules out phonon mediated pairing. They suggest an emergent Fermionic Mott insulator and doping induced superconductivity.

We use phenomenlogy, physics and microscopics to set up a relevant model to understand the same phenomenon. We have suggested that the correct model is a Bose Hubbard model, with and without doping. In various recent works, detailed band structure results and single particle information are available for twisted bilayer graphene. Our estimates of band parameters seem consistent with the band structure results, eventhough our conclusion and a Bose Hubbard model seems to be at variance.. 

The Coulomb blockade we have suggested seems to be very important for small twist angle, for example, the magic angle $\theta \approx 1.1^0$ used in recent experiments. We find that for larger angles the physics changes very quickly and our approximations do not work. Physically, when the estimated quantum confinement splitting (equation 1) exceeds interlayer tunneling, quantum confinement becomes less effective and bands are no more narrow. Large angle commensurate twists call for a different type of analysis and there may be further surprises.

In the case of perfect AA stacking, as we saw earlier, a small but finite density (4 $\times$ 10$^{12}$ /cm$^2$) of electron and hole doping is induced by a small interlayer dynamics. Do the carriers make use of RVB correlations and become superconducting ? Within our approach, we do find chiral d+id superconductivity  with a Tc $\sim$ 5 K. However, a strong electron hole attraction between nested electron and hole Fermi pockets competes with superconductivity via other instabilities. AB stacking, because of continuing absence of carriers and Fermi pockets at the Fermi level, does not have a significant superconducting instability.

While an RVB based pairing ensures a singlet superconducting order parameter, it is difficult for us to 
extract the orbital state of the Cooper pair, as it depends on details of pair scattering, the nature of states inside a puddle, valley degeneracy etc. We expect pairing involving two quantum confined states connected by time reversal symmetry. The pairs get scattered among such available time reversed pair states near the Fermi level and create a type of \textit{zero center of mass momentum state within a puddle}. Its angular momentum will in general be quenched (because of the shape of the puddle) or it may form a chiral state such as d + id. A variational Motecarlo approach along the lines of reference \cite{PathakShenoyGB} for an isolated Moire Supercell may offer some guidance. 

We would also like to point out that topological character of single particle bands close to the fermi level, will become less important for the center of mass of Cooper pairs. In some sense, Cooper pairs hide detailed topological properties of the band in the orbital relative coordinate part.

The mathematics and group theory behind Moir\'e crystallography is rich. We have only focussed on certain key aspects that are manifest in small twist angle situation.
As found by some authors, topological properties of extended states close to the Fermi level may play important role, particularly when one considers larger commensurate angles. In the present case such topological properties may play a decisive role in deciding the orbital symmetry of the spin singlet order parameter.

We also wish to remark that we have not violated Anderson's theorem on disappearance of superconductivity, with reducing size of a superconducting particle. According to Anderson theorem \cite{AndersonSmallParticleSC} superconductivity disappears when, quantum confinement induced level spacing becomes larger than the energy scale of attraction. As local energy scale of attraction J is large compared to quantum confinement induced level spacing, we do not violate Anderson theorem. 

A final remark, about the possibility of high Tc superconductivity in graphitic systems is in order. Even though twisted graphene, an ultra clean system, shows a low Tc superconductivity via a very small self doping, it should encourage a search for high Tc superconductivity \cite{Volovik1}. A prediced high Tc superconductivity in optimally doped graphene/graphite system \cite{GBMgB2} remains to be seen experimentally. It should be pointed out that there are signals for Elusive and Ustable Superconductors \cite{Kopelevich}, with high Tc's and even reaching room temperature scales in graphitic systems \cite{Esquinazi}. 

\textbf{Acknowledgement:} I thank R. Ganesh, S. Hassan, A. Jafari and M.S. Laad for discussion and comments. I am grateful to Science and Engineering Research Board (SERB, India) for award of a National Fellowship. This work, partly performed at the Perimeter Institute for Theoretical Physics, Waterloo, Canada is supported by the Government of Canada through Industry Canada and by the Province of Ontario through the Ministry of Research and Innovation.

\section*{Supplement}

\section*{Electron Correlation and Singlet-Tripet Splitting of Excited States: Benzene to Graphene}

It is known that electronic properties of p-$\pi$ bonded benzene molecule is well described by resonating valence bond theory introduced by Pauling \cite{PaulingBook}. Deby-Huckel one electron theory (filling of molecular orbitals) is less succesful. As an example, a large splitting $\sim$ 2 eV between the lowest triplet and the lowest singlet excited states, seen experimentally in benzene, is missed by Debye-Huckel theory. Electron correlation effects contained in RVB theory explains this large splitting. 

Going beyond benzene, in his study of graphite, Pauling described \cite{PaulingBook} 2d graphene using an RVB wave function on a honeycomb lattice. After Anderson's work \cite{PWARVB1973} it became clear that Pauling's RVB wave function for graphene actually describes a Mott insulating spin liquid with a strong singlet pairing, rather than a semimetal. Pauling's RVB wave function did not contain important polar (charge) fluctuations needed to describe a semimetallic state of graphene. 

With hind sight one can say that Pauling was ahead of his time (not unusual) and exposed to us a hidden singlet pairing correlation, which may become valuable later. Fermi liquid theory (a theory valid at low energies) for semimetallic graphene, also misses the off shell RVB correlations. Effects of RVB correlations appear only as renormalized Fermi liquid parameters.

RVB wave functions describe strongly correlated electronic states, where charge degrees of freedom are frozen, in the sense that there is a finite energy gap to create charged excitations. In this background of \textit{frozen charges}, pairs of spins form spin singlets and resonate. We get a quantum spin liquid. An RVB state can not be written as a Slater determinant - it defies an independent electron description. A popular, short range RVB state is obtained by a coherent quantum superposition of all configuration of singlet paired nearest neighbors in a given lattice. Equation (8) to be discussed in the next section is an example of RVB wave function, when $\alpha$ = 0. It is characterized by a (valence bond) singlet pair function $\phi_{ij}$.

Akbar Jafari and present author found \cite{GBAkbarSpin1PRL} that the singlet-triplet splitting found in molecular benzene survives in the extended semimetallic graphene, in the form of an emergent spin-1 collective mode branch, This mode lies in the \textit{window of particle-hole continuum}, and occupies a large part of the Brilluouin zone, without getting Landau damped. We also found that the collective mode formation keeps the Fermi liquid character of semimetallic graphene intact. That is, low energy physics of graphene is captured well by Fermi liquid theory of Dirac semimetal.

Nevertheless, we interpreted the existence of a spin-1 collective mode as evidence from manybody theory for presence of Pauling's RVB physics in the ground state of semimetallic graphene. Singlet correlations present in the ground state enable formation of excited triplet branch, obtained by breaking of existing singlet bonds.\\

\section*{Theory of High Tc Supercondudctivity in Doped Graphene: A Summary}

In this section we review our early theory of high Tc superconductivity in doped graphene like systems, arising from even a moderate electron-electron repulsion. At the heart of our suggestion is the low dimensionality of graphene. It is known from Lieb-Wu solution \cite{LiebWu} that in 1d repulsive Hubbard model at half filling, any small repulsive U gets renormalized upwards (via repeated scattering in 1d), resulting in a Mott Hubbard gap and a finite effective singlet spin-spin coupling J between two neighboring sites. Both are non perturbative effects.
 
We suggested \cite{GBMgB2} that even a moderate repulsive Hubbard U present in 2d graphene, stablizes a pairwise entanglement (a finite dynamical spin spin singlet coupliing J) between electron spins at neighboring sites, via repeated Hubbard U scattering in spin singlet channel. Unlike 1d, a small repulsive U does not cause Mott localization however. More importantly an emergent well developed singlet correlation in neutral graphene remains latent and does not result in superconductivity, because of absence of carriers at the Fermi level in neutral graphene. 

Since Fermi liquid theory works remarkably well for neutral graphene, RVB correlations tend to be ignored in general in the literature. Can we afford to ignore hidden RVB correlations always ?  Inspired by the discovery of superconductivity in MgB$_2$ (containing graphene like charged B$^-$ honeycomb layers and AA stacking, intercalated by a triangular insulating lattice of Mg$^{++}$ ions), the present author addressed the above question and presented an effective theory \cite{GBMgB2}, where RVB correlations manifest as superconductivity via internal doping (electron transfer from $\sigma$ band to $\pi^*$ band) in a graphite like structures.

We suggested that electron correlation induced bond singlet correlation which are otherwise dormant start showing up at high energy scales, or when perturbed. In particular, when we perturb the system by adding carriers and move away from charge neutral point of graphene, high Tc superconductivity emerges for a range of optimal doping, via electron correlation induced singlet pairing. \textit{That is, two added carriers have a strong nearest neighbor singlet pairing attraction via RVB physics},
resulting in superconductivity, with Tc decided by the amount of doping.

We view the induced nearest neighbor singlet coupling between two added carriers as an on shell pheonomenon in the following sense. Two added electrons, for example, have more phase space for scattering (entire $\pi^*$ band). Further, these two electrons are quantum mechanically indistinguishable from electrons of the dense $\pi$ electron fluid. Thus added electrons also feel RVB correlations and are able to bring it on shell in their dynamics.

Our original theory introduced an effective Hamiltonian, which had a nearest neighbor singlet pairing pseudopotential J for added carriers. It is a phenomenological coupling constant that took into account onsite Coulomb repulsion induced nearest neighbor spin spinglet pairing. When we apply our theory to benzene, required J value is large $\sim$ 3 eV, reflecting strong singlet correlations present in the ground state. For graphene this J is reduced because of $\pi\pi^*$ band formation. However, even the reduced J $\sim$ 1 to 2 eV, in our estimate \cite{GBAkbarSpin1PRL}, is still large, leading to interesting prospects for high Tc superconductivity.

We studied \cite{GBMgB2} the above effective Hamiltonian using mean field theory. Our theory correctly captures absence of superconductivity in neutral graphene and prediced high Tc superconductivity on doping. Our approach was pursued by Doniach and Black-Schaffer \cite{DoniachAnnica}, who discovered a chiral mean field solution with a d + id symmetry for the superconducting order parameter, and quantified our high scale of mean field Tc at optimal doping. 

Encouraged by the above development, and in the wake of a growing interest in graphene, Pathak, Shenoy and the present author \cite{PathakShenoyGB}, in 2010, went beyond mean field theory, and directly analysed the 2d repulsive Hubbard model on a honeycomb lattice, with parameters chosen for graphene (t $\approx$ 3 eV and U $\approx$ 6 eV ). We used a state of the art variational Montecarlo approach and studied the following variational wave function at various dopings:

\begin{equation}
|\alpha,~\Delta \rangle \equiv \prod_i (1 - \alpha ~n_{i\uparrow}n_{i\downarrow}) ~[\sum_{ij} \phi_{ij} b^\dagger_{ij}]^{N/2}~ |0\rangle
\end{equation}
Here $n_{i\uparrow},n_{i\downarrow}$ are electron number operators and  $b^\dagger_{ij} \equiv
\frac{1}{\sqrt{2}} (c^\dagger_{i\uparrow}c^\dagger_{j\downarrow} - c^\dagger_{i\downarrow} c^\dagger_{j\uparrow})$ is the singlet electron pair operator. N is the number of electrons.

Our wave function contains two variational parameters $\alpha$ and $\Delta$. $\alpha$ is the Gutzwiller parameter. $\Delta$ is spin singlet d+id pairing chiral order parameter. It was introduced via the Cooper pair function $\phi_{ij}(\Delta)$. When $\Delta$ = 0 and $\alpha$ = 0 equation (8) becomes the non-interacting ground state.

We minmized energy of the repulsive Hubbard model with respect to $\alpha$ and $\Delta$. For zero doping we found an optimal solution $\Delta$ = 0 and established a partially Gutzwiller projected non superconducting Dirac semimetal as the ground state. We calculated spatial dependence of two particle singlet pair correlation for a range of dopings and extracted coherence length. We intepreted our result using a weak coupling BCS formula and estimated the largest superconducting gap to be $\sim$ 50 - 100 meV, and an optimal doping of 15 \%. Corresponding Tc, in our variational approach, which took into account quantum fluctuations, was encouragingly high in the range of 200 K. Subsequent works have gone further and have have presented interesting analysis \cite{Nandkishore,Thomale,Annica2014,ZGu,DungHai}

\end{document}